\documentclass[twocolumn,showpacs,preprintnumbers,amsmath,amssymb]{revtex4}
\usepackage{epsfig}
\newcommand{\bm}[1]{ \mbox{\boldmath $#1$}  }

\begin{document}

\title{Origin of three-body resonances }

\author{E. Garrido}
\affiliation{Instituto de Estructura de la Materia, CSIC, Serrano 123, E-28006
Madrid, Spain}
\author{D.V. Fedorov}
\author{A.S. Jensen}
\affiliation{Department of Physics and Astronomy, University of Aarhus, 
DK-8000 Aarhus C, Denmark}
\date{\today}

\begin{abstract}
We expose the relation between the properties of the three-body
continuum states and their two-body subsystems. These properties
refer to their bound and virtual states and resonances, all
defined as poles of the $S$-matrix. For one infinitely heavy core and
two non-interacting light particles, the complex energies of the
three-body poles are the sum of the two two-body complex
pole-energies.  These generic
relations are modified by center-of-mass effects which alone can
produce a Borromean system.  We show how the three-body states evolve
in $^6$He, $^6$Li, and $^6$Be when the nucleon-nucleon interaction
is continuously  switched on.  The schematic model is able to reproduce
the main properties in their spectra.  Realistic calculations for
these nuclei are shown in detail for comparison. The implications of a
core with non-zero spin are investigated and illustrated for $^{17}$Ne
($^{15}$O+$p$+$p$). Dimensionless units allow predictions for systems
of different scales.
\end{abstract}


\maketitle

\section{Introduction}

The three-body problem has a long history from macroscopic celestial
classical mechanics, e.g. sun-earth-moon \cite{gut98} to microscopic
quantum mechanics, e.g. the helium atom \cite{tan01}, three nucleons
\cite{kie94} or three quarks \cite{ric92}. The modern treatment was
boosted by the formulation of the Faddeev equations \cite{fad61}
originally aimed at scattering problems, but also successfully applied
for bound states \cite{car03,nie01}. The more recent interest in bound
state halo structures and Borromean systems are by now fairly well
understood in terms of the basic two-body interactions \cite{jen04}.
The success is at least indisputable for three-body systems with only
one or a few bound states.

On the other hand, the corresponding three-body properties for
positive energies (energies above breakup) are much less established 
although studied
thoroughly for both short and long range interactions
\cite{glo96,tan01}. Contributions from both short and long-range 
interactions make computations numerically difficult.  The three-body
Coulomb problem itself is still considered unsolved
\cite{lin95,res99,alt04} and three-body resonances for strongly
interacting particles are still debated
\cite{shu00,myo01,bet03,mic02,del00}.  This is unfortunate, since the
continuum structure often form the basis in descriptions of the
dynamic behavior of a given system.

Crucial properties of the continuum are revealed by information about
the resonances and virtual states defined as poles of the $S$-matrix.
Together with the discrete set of bound negative-energy states 
we also have a discrete set of unbound complex-energy states with their 
corresponding wave functions. For completeness also the continuous non-singular
background states are needed, but the singularities of the scattering
matrix are very often decisive.  Important examples using different
methods within few-body physics are astrophysical reaction rates
\cite{gor95}, adiabatic reaction processes arising at low energies or
for large impact parameters \cite{gar01}, three-body decays
\cite{gar05}, three-body resonances for Faddeev operators
\cite{kol01}, for nucleons \cite{wit99}, for electrons and positrons
\cite{pap02} and four-body nuclear continuum states \cite{laz04}.  This 
list could be extended.

Continuum structure is in general more difficult than bound state
problems although various methods have been designed to overcome the
technical problems at least for resonances, see for example
\cite{glo96,myo01,mic02,bet03,gar03,ho83}. An understanding of the 
generic origin of $S$-matrix poles would be tremendously helpful
especially if recognizable traces of a well-structured origin are left
in the realistic spectra.  No doubt this would allow easier
interpretation of complicated numerical results, allowed design of
better methods, and indicate which effects to look for in different
contexts.

The purpose of this work is to investigate how the two-body interactions 
determine the three-body continuum structure. We focus on three-body
continuum states where none of the two-body subsystems is bound (Borromean
systems). When two-body bound states are possible, different three-body
structures can  appear, as for instance, unbound three-body states with 
negative energy. However, these systems can very often be
understood as two-body systems made by a two-body bound state and the
remaining third particle in the continuum.  With the
help of the hyperspheric adiabatic expansion method we determine which
three-body bound or virtual states and resonances result from given
sets of corresponding two-body properties. Taking a simple system as
starting point, it is then possible to observe how the three-body
states evolve in the complex energy plane when more and more realistic
features are incorporated into the calculation.

In section II we give details about the complex scaled adiabatic
expansion methods. In section III the schematic system of an
infinitely heavy core and two mutually non-interacting light particles
is described analytically, while in section IV the general properties
of the system are described after numerical studies of specific
systems. In section V we relate the spectra from the schematic model
with those obtained in realistic calculations for systems with zero
($^6$He, $^6$Li, $^6$Be) and non-zero core-spin ($^{17}$Ne).  We finish
in section VI with some qualitative estimates for other systems and in
section VII we give a short summary and the conclusions.

\section{The complex scaled hyperspheric adiabatic expansion method}

To describe a three-body system we use the standard coordinates:
\begin{eqnarray}
\bm{x}_i&=&\sqrt{\frac{m_j m_k}{m (m_j+m_k)}}(\bm{r}_j-\bm{r}_k) \;,
 \label{eq1} \\ \nonumber 
\bm{y}_i&=&\sqrt{\frac{m_i(m_j+m_k)}{m (m_i+m_j+m_k)}}
  \left(\bm{r}_i-\frac{m_j\bm{r}_j+m_k\bm{r}_k}{m_j+m_k}\right),
\end{eqnarray}
where $m_i$, $m_j$, and $m_k$ are the masses of the three particles
and $m$ is an arbitrary normalization mass.  From the Jacobi
coordinates we define the hyperspheric coordinates
$\{\rho,\Omega_{i}\}$, where $\{\Omega_i\} =
\{\alpha_i,\Omega_{x_i},\Omega_{y_i}\}$, $\rho$=$\sqrt{x_i^2+y_i^2}$,
$\alpha_i$=$\arctan(x_i/y_i)$, and $\Omega_{x_i}$ and $\Omega_{y_i}$
give the directions of $\bm{x}_i$ and $\bm{y}_i$, respectively.

The three-body wave function is written as a sum of three components
$\psi^{(i)}(\bm{x}_i,\bm{y}_i)$, each corresponding to one of the
three possible sets of Jacobi coordinates \cite{nie01}. These three
components satisfy the three Faddeev equations
\begin{eqnarray}
 & & (T-E) \psi^{(i)}(\bm{x}_i,\bm{y}_i) \label{eq2} + V_{jk}(x_i)\\ & &
 \times \left(
\psi^{(i)}(\bm{x}_i,\bm{y}_i)+\psi^{(j)}(\bm{x}_j,\bm{y}_j)+
\psi^{(k)}(\bm{x}_k,\bm{y}_k)
\right)=0, \nonumber
\end{eqnarray}
where $T$ is the kinetic energy operator, $V_{jk}(x_i)$ is the
two-body interaction between particles $j$ and $k$, and $E$ is the
three-body energy.  Here $(i,j,k)$ is a cyclic permutation of $(1,2,3)$.

We employ the hyperspheric coordinates to expand each component
$\psi^{(i)}(\bm{x}_i,\bm{y}_i)$ in terms of a complete set of angular
functions $\phi_n^{(i)}$
\begin{equation}
\psi^{(i)}_{n_0}(\bm{x}_i,\bm{y}_i)
   =\frac{1}{\rho^{5/2}} \sum_n f^{(n_0)}_n(\rho) \phi_n^{(i)}(\rho,\Omega_i),
\label{eq3}
\end{equation}
where the additional index $n_0$ labels different solutions we later
on want to consider. Usually the expansion (\ref{eq3}) converges
rather fast, and only a few terms (typically no more than three) are
needed.

By rewriting Eq.(\ref{eq2}) in terms of the hyperspheric coordinates,
and inserting the expansions in Eq.(\ref{eq3}) we separate the Faddeev
equations into angular and radial parts:
\begin{eqnarray}
(\hat{\Lambda}^2-\lambda_n(\rho)) \phi_n^{(i)} \hspace{5.0cm} \label{eq4} \\ 
 \nonumber   - \frac{2 m \rho^2}{\hbar^2} V_{jk}^{(i)}(x_i)
\left( \phi_n^{(i)} + \phi_n^{(j)}  + \phi_n^{(k)}   \right) = 0  \; ,  
 \nonumber      \\  
\left[ -\frac{d^2}{d\rho^2} - \frac{2mE}{\hbar^2}+ \frac{1}{\rho^2}
\left( \lambda_n(\rho)+\frac{15}{4} \right) \right] f^{(n_0)}_n(\rho) 
 = \nonumber  \\ 
 \sum_{n'} \left( 2 P_{n n'} \frac{d}{d\rho} + Q_{n n'} \right)
 f_{n'}^{(n_0)}(\rho) \;,
\label{eq5}
\end{eqnarray}
where $\hat{\Lambda}^2$ is a hyperangular operator which together with
expressions for the functions $P_{n n'}(\rho)$ and $Q_{n n'}(\rho)$
can be found in \cite{nie01}.  The complete set of angular functions
$\phi_n^{(i)}$ in the expansion (\ref{eq3}) is precisely the
eigenfunctions of the angular part of the Faddeev equations.  The
index $n$ labels the the corresponding eigenvalue $\lambda_n$, which
enters in the coupled set of radial equations (\ref{eq5}) as effective
potentials.  The index $n_0$ is related to the boundary condition for
continuum wave functions, which for short-range interactions and no
two-body bound states has the asymptotics given by \cite{nie01}:
\begin{eqnarray}
&& f_{n}^{(n_0)}(\kappa \rho) \longrightarrow \label{eq6} \\ \nonumber 
&& \sqrt{\kappa\rho}\left[
H_{K+2}^{(2)}(\kappa\rho)\delta_n^{n_0}+S_n^{n_0}(\kappa)
H_{K+2}^{(1)}(\kappa\rho) \right]  ,
\end{eqnarray}
where $n_0$ then labels the incoming channel, the $S$-matrix
$S_n^{n_0}$ is a factor depending on the complex momentum $\kappa$
related to the complex energy $E$ by $\kappa$=$\sqrt{2mE/\hbar^2}$,
and $H_{K}^{(1)}$ and $H_{K}^{(2)}$ are Hankel functions of first and
second kind.  The hypermomentum $K$ is given by the asymptotic
behaviour of the angular eigenvalue $\lambda_n(\rho)$ that approaches
$K(K+4)$ \cite{nie01}.

For three-body bound states $f_{n}^{(n_0)}$ must fall off
exponentially at large distances.  Then only one Hankel function is
present and the $S$-matrix $S_n^{n_0}$ has a pole for the imaginary
momentum $\kappa$ ($\kappa=i|\kappa|$) and the energy $E$ is negative.  
Poles of the
$S$-matrix in the lower half of the complex momentum plane correspond
to three-body resonances, and their asymptotic radial wave function is
given only by the $H_{K+2}^{(1)}$ part in Eq.(\ref{eq6}).
Asymptotically the Hankel function $H_{K+2}^{(2)}$ vanishes
exponentially like $e^{-|\kappa|\rho}$, while $H_{K+2}^{(1)}$ grows
like $e^{|\kappa|\rho}$.  This means that the radial coefficients of
the continuum wave functions are dominated by the $H_{K+2}^{(1)}$
part. Therefore, it is very difficult to distinguish the resonance
wave function, where only the $H_{K+2}^{(1)}$ term is present in the
asymptotics, from an ordinary continuum wave function.

This problem is solved by applying the complex scaling method 
\cite{ho83,cso94,cso94b,aoy95,aoy95b}, where
the Jacobi coordinates $x_i$ and $y_i$ are rotated into the complex
plane by an arbitrary angle $\theta$ ($x_i\rightarrow x_i
e^{i\theta}$, $y_i\rightarrow y_i e^{i\theta}$).  This means that only
the hyperradius $\rho$ is rotated ($\rho\rightarrow \rho
e^{i\theta}$), while the five hyperangles $\{\Omega_i\}$ remain
unchanged.  After this rotation the radial wave functions of the
resonances behave asymptotically like
\begin{eqnarray}
f_{n}^{(n_0)}(\kappa\rho e^{i\theta})  \rightarrow 
\sqrt{\rho} H_{K+2}^{(1)}(|\kappa| \rho e^{i(\theta-\theta_R)})
   \label{eq7} \\
 \rightarrow 
     e^{-|\kappa|\rho \sin{(\theta-\theta_R)}} e^{i\left(|\kappa|\rho
\cos{(\theta-\theta_R)}-K\pi/2+3\pi/4 \right)}, \nonumber 
\end{eqnarray}
where $\theta_R$ is the argument of the complex momentum $\kappa$
($\kappa=|\kappa|e^{-i\theta_R}$). From (\ref{eq7}) we observe that
when $\theta>\theta_R$ the radial wave function falls off
exponentially, exactly as a bound state. Continuum wave functions are
dominated in the asymptotics by the $H_{K+2}^{(2)}(|\kappa|
\rho e^{i(\theta-\theta_R)})$ term, that 
diverges exponentially when $\theta>\theta_R$.  Thus, after complex
scaling, resonances can be easily distinguished from ordinary
continuum states, and furthermore, the numerical technique used to
compute bound states can be used for resonances. In particular, after
solving the complex scaled equations (\ref{eq4}) and (\ref{eq5}) with
the boundary condition (\ref{eq7}) three-body resonances and bound
states are simultaneously obtained. Solving Eq.(\ref{eq5}) with a box
boundary condition ($f(\rho_{max})$=0, $\rho_{max}$ being a large
value of the hyperradius) discretizes the continuum spectrum, and the
continuum states are rotated by an angle $\theta$ in the momentum
plane and $2\theta$ in the energy plane \cite{ho83}. The box boundary
condition is often enough to obtain accurate bound state and resonance
solutions.

\section{Two independent subsystems}

We analyze in this section the schematic model of an infinitely heavy
core and two light particles. These two mutually non-interacting light
particles interact with the core via short-range interactions.  We
shall first assume that these three particles have zero spin and then
later generalize to non-zero particle spins.

\subsection{Particles without spin}

\begin{figure}
\begin{center}
\epsfig{file=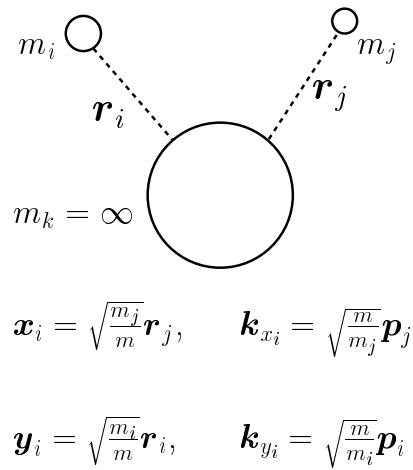,scale=0.75}
\end{center}
\caption[]{Sketch of the schematic three-body system. Particles $i$ and 
$j$ do not interact with each other. Particle $k$ is infinitely
heavy. One set of Jacobi coordinates and their corresponding conjugate
momenta are given in the lower part. }
\label{fig1}
\end{figure}

From the Jacobi coordinates (\ref{eq1}) we can construct the 
corresponding conjugated momenta:
\begin{small}
\begin{eqnarray}
\bm{k}_{x_i}&\!\!\!\!=
   &\!\!\!\!\!\sqrt{\frac{m m_j m_k}{m_j+m_k}}
   \left(\frac{\bm{p}_j}{m_j}-\frac{\bm{p}_k}{m_k}\right) \; ,
\label{eq8} \\
\bm{k}_{y_i}&\!\!\!\!=
  &\!\!\!\!\!\sqrt{\frac{m m_i(m_j+m_k)}{(m_i+m_j+m_k)}}
  \left(\frac{\bm{p}_i}{m_i}- \frac{\bm{p}_j+\bm{p}_k}{m_j+m_k}\right).
\nonumber
\end{eqnarray}
\end{small}
In precise analogy to the hyperradial coordinates, we now introduce
the hypermomentum coordinates $\{\kappa,\Omega_{\kappa_i}\}$ defined
as $\{\Omega_{\kappa_i}\}= \{\alpha_{\kappa_i},\Omega_{k_{x_i}},
\Omega_{k_{y_i}}\}$,  $\kappa=\sqrt{k_{x_i}^2+k_{y_i}^2}$,
$\alpha_{\kappa_i}=\arctan(k_{x_i}/k_{y_i})$. The three-body momentum
$\kappa$ defined earlier is independent of the Jacobi coordinate
system labeled $i$. The directions of $\bm{k}_{x_i}$ and
$\bm{k}_{y_i}$ are described by $\Omega_{k_{x_i}}$ and
$\Omega_{k_{y_i}}$, respectively.

When one of the three spinless particles has infinite mass (let us
take $m_k$=$\infty$), Eqs.(\ref{eq1}) and (\ref{eq8}) simplify, in the
three-body center-of-mass system where $\bm{r}_k$=0 and
$\bm{p}_i$+$\bm{p}_j$+$\bm{p}_k$=0, to the form in Fig.\ref{fig1}.
Therefore, in two of the three possible sets of Jacobi coordinates,
the coordinates are simply proportional to the distances between the
infinitely heavy core and each of the two light particles.  If the
two-body interactions depend only on the distance between particles,
the three-body Hamiltonian $H$ can be written as $H=H_{ki}+H_{kj}$
(see Fig.\ref{fig1}) where $H_{ki}$ and $H_{kj}$ are the two-body
Hamiltonians describing the corresponding two-body systems.  From this
separability, the three-body bound state energies are obviously given
by the sum of the corresponding two-body bound state energies.

For three-body resonances and virtual states the same
happens. Three-body resonances show up at energies equal to the sum of
the two-body resonance energies of the $ki$ and $kj$ subsystems.  From
the separability of the Hamiltonian this result may be accepted as
obvious. However, the only trivial deduction using the separability is
that the three-body energy is given by the sum of the two two-body
energies.  This does not necessarily imply that a three-body energy
equal to the sum of two two-body resonance energies must correspond to
a three-body resonance.  Bound states have discrete energies, and the
split of a given three-body bound state energy into two two-body
energies can be made only in specific ways, since the two-body
energies can have only the specific values corresponding to the
discrete two-body bound state spectrum.  However, resonances are
continuum states, and then a given three-body energy can be obtained
by infinitely many pairs of two-body energies. Therefore, if a given
three-body energy matches with the sum of two two-body resonance
energies, there are also infinitely many other pairs of continuum
(non-resonant) two-body energies whose sum gives the same three-body
energy.  Thus, it is not a trivial conclusion that only the particular
three-body state of matching energy corresponds to a three-body
resonance defined as a pole in the three-body $S$-matrix.

Let us then investigate this more closely.  From the separability of
the Hamiltonian it follows that the three-body wave function is the
product of the wave functions describing the $ki$ and $kj$ two-body
subsystems.  For unbound two-body systems the two-body wave function
can be expanded in partial waves, and the three-body wave
function takes then the form:
\begin{eqnarray}
\lefteqn{
\Psi^{(+)}(\bm{x},\bm{y},\bm{k}_x,\bm{k}_y)=} \label{eq10} \\ &&
 \left( \sqrt{\frac{2}{\pi}}
  \sum_{\ell_x m_{\ell_x}} i^{\ell_x} \frac{u_{\ell_x}(x,k_x)}{x k_x}
  Y_{\ell_x m_{\ell_x}}(\Omega_x)
  Y_{\ell_x m_{\ell_x}}^\star(\Omega_{k_x}) \right)
 \nonumber \\  & \times &
 \left( \sqrt{\frac{2}{\pi}}
  \sum_{\ell_y m_{\ell_y}} i^{\ell_y} \frac{u_{\ell_y}(y,k_y)}{y k_y}
  Y_{\ell_y m_{\ell_y}}(\Omega_y)
  Y_{\ell_y m_{\ell_y}}^\star(\Omega_{k_y}) \right)
\nonumber
\end{eqnarray}
where $(\bm{x},\bm{y})$ refer to the Jacobi system in Fig.\ref{fig1}
and the momenta of the corresponding subsystems are $\bm{k}_x$ and
$\bm{k}_y$. The spherical harmonics and the two-body radial functions
are denoted $Y_{\ell m_{\ell}}$ and $u_{\ell}$.

Following \cite{gar01}, also the continuum wave function for a
three-body system of spinless particles can be written as:
\begin{eqnarray}
\lefteqn{
\Psi^{(+)}(\bm{x},\bm{y},\bm{k}_x,\bm{k}_y) =} \label{eq9} \\ & &
\!\!\!\!\!\sum_{K L M \ell_x \ell_y} \!\!\!\!\!
Y^{(\ell_x \ell_y)*}_{K L M}(\Omega_{\kappa})
\sum_{K^\prime L^\prime M^\prime \ell_x^\prime \ell_y^\prime}
\!\!\!\!\!
f^{(K^\prime \ell_x^\prime \ell_y^\prime L^\prime)}_{K \ell_x \ell_y L}
(\kappa \rho)
Y^{(\ell_x^\prime \ell_y^\prime)}_{K^\prime L^\prime M^\prime}(\Omega)
 \nonumber
\end{eqnarray}
where $(\bm{x},\bm{y})$, $(\bm{k}_x,\bm{k}_y)$, $\Omega$ and
$\Omega_{\kappa}$ could correspond to any of the three possible sets
of Jacobi systems. We omitted the index $i$. The hyperradial
functions, $f^{(K^\prime \ell_x^\prime \ell_y^\prime L^\prime)}_{K
\ell_x \ell_y L}$, are solutions to Eq.(\ref{eq5}) where we now specified 
the indices $n$ and $n_0$ and explicitly included the factor $\kappa$
in the argument.  The hyperspheric harmonics $Y^{(\ell_x \ell_y)}_{K L
M}(\Omega)$ can for instance be found in \cite{nie01}.  Using the
orthogonality of the hyperspheric harmonics, we immediately have
\begin{eqnarray}
\lefteqn{
f^{(K^\prime \ell_x^\prime \ell_y^\prime L^\prime)}_{K \ell_x \ell_y L}
(\kappa \rho) = } \label{eq11} \\ & &
\!\!\int d\Omega \!\! \int d\Omega_{\kappa} 
 \Psi^{(+)}(\bm{x},\bm{y},\bm{k}_x,\bm{k}_y)
 Y^{(\ell_x \ell_y)}_{K L M}(\Omega_{\kappa})
 Y^{(\ell_x^\prime \ell_y^\prime)*}_{K^\prime L^\prime M^\prime}(\Omega)
\nonumber
\end{eqnarray}
and by inserting Eq.(\ref{eq10}) into (\ref{eq11}) we get
\begin{eqnarray}
&&
\!\!\!\!\!\!\!
f^{(K^\prime \ell_x^\prime \ell_y^\prime L^\prime)}_{K \ell_x \ell_y L}
(\kappa \rho) =\delta_{\ell_x \ell_x^\prime} \delta_{\ell_y \ell_y^\prime} 
 \delta_{L L^\prime}
 \frac{2}{\pi} i^{\ell_x+\ell_y} N_K^{\ell_x \ell_y} 
 N_{K^\prime}^{\ell_x \ell_y}  \nonumber \\
& & \!\!\!\!\!\!\!\times
\int_0^{\pi/2} \!\!\!\!d\alpha_\kappa (\sin{\alpha_\kappa})^{\ell_x+2}
(\cos{\alpha_\kappa})^{\ell_y+2} 
P_n^{(\ell_x+\frac{1}{2},\ell_y+\frac{1}{2})}(\cos{2\alpha_\kappa})
\nonumber \\
& & \!\!\!\!\!\!\!\times 
\int_0^{\pi/2} \!\!\!\!d\alpha (\sin{\alpha})^{\ell_x+2}
(\cos{\alpha})^{\ell_y+2} 
P_{n^\prime}^{(\ell_x+\frac{1}{2},\ell_y+\frac{1}{2})}(\cos{2\alpha})
\nonumber \\ & & \hspace*{3.0cm}
\times \frac{u_{\ell_x}(x,k_x)}{x k_x} \frac{u_{\ell_y}(y,k_y)}{y k_y} \; ,
\label{eq12}
\end{eqnarray}
where $N_K^{\ell_x \ell_y}$ is the normalization constant appearing
when expressing the hyperspheric harmonics in terms of the spherical
harmonics and the Jacobi polynomials \cite{nie01}.  Due to the delta
functions, the three-body hyperradial functions in Eq.(\ref{eq12}) are
block-wise diagonal and each block matrix can, for given $(\ell_x,
\ell_y, L)$, be labeled by the indexes $KK^\prime$.  Note also that 
the wave function does not depend on the total orbital angular
momentum $L$.  All $L$-values allowed by coupling are degenerate.

In the limit of zero interactions between the core and the light particles
the two-body radial wave functions $u_\ell(x,k)/xk$  become the spherical
Bessel function $j_\ell(xk)$, and Eq.(\ref{eq12}) becomes:
\begin{equation}
f^{(K^\prime \ell_x^\prime \ell_y^\prime L^\prime)}_{K \ell_x \ell_y L}
(\kappa \rho)
=\delta_{\ell_x \ell_x^\prime} \delta_{\ell_y \ell_y^\prime} \delta_{L L^\prime}
 \delta_{K K^\prime} i^K \frac{J_{K+2}(\kappa\rho)}{(\kappa\rho)^2},
\label{eq13}
\end{equation}
where $J_{K+2}$ is a Bessel function of the first kind. Inserting
(\ref{eq13}) into (\ref{eq9}) we then obtain the three-body wave
function
\begin{eqnarray}
& & \Psi^{(+)}(\bm{x},\bm{y},\bm{k}_x,\bm{k}_y)= \nonumber \\
& & \sum_{K \ell_x \ell_y L M} i^K \frac{J_{K+2}(\kappa\rho)}{(\kappa\rho)^2} 
Y^{(\ell_x \ell_y)*}_{K L M}(\Omega_\kappa)
Y^{\ell_x \ell_y}_{K L M}(\Omega)   \\
& & \hspace*{4cm} \equiv  
\frac{1}{(2\pi)^3} e^{i\left(\bm{k}_x\cdot \bm{x}+\bm{k}_y\cdot \bm{y}\right)}
\nonumber
\end{eqnarray}
and the plane waves are recovered from the hyperradial partial wave expansion.

The large-distance asymptotic behaviour of the three-body radial wave
functions, Eq.(\ref{eq12}), can be obtained from the asymptotics of
the two-body wave functions, that for the case in which none of the 
two-body core-particle subsystems is bound has the form:
\begin{equation}
\frac{u_\ell(x,k)}{x k} \rightarrow \frac{1}{2}\left(
h_\ell^{(2)}(xk)+s_\ell(k)h_\ell^{(1)}(xk)
\right) \; ,
\label{eq15}
\end{equation}
where $h_\ell^{(i)}$ are the corresponding two-body Hankel functions
of first and second kind, and $s_\ell(k)$ is the two-body $S$-matrix.
Inserting Eq.(\ref{eq15}) into (\ref{eq12}) and using the expressions
given in the appendix, we obtain
\begin{eqnarray}
&&f^{(K^\prime)}_{K}(\kappa \rho) \longrightarrow \label{eq16} \\ \nonumber   
&& \frac{i^{K^\prime}}{(\kappa\rho)^2}\left[
H_{K+2}^{(2)}(\kappa\rho)\delta_{K}^{K^\prime}+S_{K}^{K^\prime}
(\kappa)
H_{K+2}^{(1)}(\kappa\rho) \right] , \;
\end{eqnarray}
where we omitted the unimportant block indexes $(\ell_x,\ell_y,L)$.  Then
Eq.(\ref{eq16}) is identical to Eq.(\ref{eq6}) after division by the
phase factor $1/\rho^{5/2}$ in Eq.(\ref{eq3}).  The three-body
$S$-matrix takes the form:
\begin{equation}
S_K^{K^\prime}(\kappa)= \frac{2
\int_0^{\pi/2} d\alpha_{\kappa}\Phi_{n n^\prime}^{(\ell_x \ell_y)}
(\alpha_\kappa) s_{\ell_x}(k_x) s_{\ell_y}(k_y) }{ 
\int_0^{\pi/2} d\alpha_{\kappa}\Phi_{n n^\prime}^{(\ell_x \ell_y)}
(\alpha_\kappa) (s_{\ell_x}(k_x)+s_{\ell_y}(k_y))}
\label{eq17}
\end{equation}
where
\begin{eqnarray}
\Phi_{n n^\prime}^{(\ell_x \ell_y)}(\alpha_\kappa) & = &  
(\sin{\alpha_\kappa})^{2\ell_x+2}
(\cos{\alpha_\kappa})^{2\ell_y+2} \\ & & \hspace*{-1.5cm}\times
P_n^{(\ell_x+\frac{1}{2},\ell_y+\frac{1}{2})}(\cos{2\alpha_\kappa})
P_{n^\prime}^{(\ell_x+\frac{1}{2},\ell_y+\frac{1}{2})}(\cos{2\alpha_\kappa}),
\nonumber
\end{eqnarray}
is a smooth function of $\alpha_\kappa$ given in terms of the Jacobi
functions $P_n^{(\ell_x+\frac{1}{2},\ell_y+\frac{1}{2})}$ with $K= 2n+
\ell_x + \ell_y$ and $K^\prime= 2n^\prime + \ell_x + \ell_y$.

The integrands in Eq.(\ref{eq17}) contain the two-body $S$-matrices
depending on corresponding momenta.  The integration variable
maintains the total energy of the two subsystems while varying the
ratio of their momenta.  Assume that one pole of one of the subsystems
is a first order pole which does not coincide with any pole of the
other subsystem, then the integrations in Eq.(\ref{eq17}) across the
pole give a smooth function without trace of the pole. However, if
two poles from different subsystems exist simultaneously the product of 
two-body $S$-matrices in the numerator of Eq.(\ref{eq17}) results in a pole
term of second order, which after integration across the pole, leaves
a first order pole in the three-body $S$-matrix. The energy of this
three-body pole corresponds then precisely to the sum of the energies
of the two two-body poles.

If one pole is of higher order, and not coinciding with a pole from
the other subsystem, it would survive the integration in precisely in
the same way in numerator and denominator. Thus no trace is left in
the three-body $S$-matrix. A three-body pole only arises when two
two-body poles belonging to different subsystems exist simultaneously.  
Combining
any order of coinciding poles from the two subsystems is then seen to
produce a three-body pole of an order equal to the smallest order of
the two-body poles involved. This proves that the three-body poles
appear if and only if the energy is equal to the sum of energies of
two two-body poles from different subsystems.  The smallest order of
the two coinciding poles reappears in the three-body pole.  

In case of having a two-body bound state, the corresponding two-body asymptotics
is not given by Eq.(\ref{eq15}) but by a falling off exponential $e^{-kx}$, 
where
$k$ is determined by the two-body binding energy. The three-body properties
are then directly dictated by the two-body $S$-matrix describing the
remaining core-particle subsystem. If this two-body $S$-matrix has a pole
corresponding to a two-body bound state, the three-body system is obviously
bound with a binding energy equal to the sum of the two two-body binding
energies. If the two-body $S$-matrix has a pole corresponding to a two-body
resonance (or virtual state), the three-body system has then a resonance
(or virtual state) that actually corresponds to a resonance (or virtual state)
of the two-body system made by the bound two-body state and the remaining third
particle.

At this point it is also important to mention that we are identifying poles
of the $S$-matrix in the fourth quadrant of the energy plane and resonances. 
However, strictly speaking, for a pole to be considered a resonance it is 
required that its width is clearly smaller than the energy separation between 
resonances. In this way resonances are the poles of the $S$-matrix close 
enough to the real energy axis. It can certainly happen that when two complex 
energies corresponding to two singularities in the $S$-matrix are added, the 
final energy can be close to the negative imaginary axis, or even in the
third quadrant of the energy plane. These singularities should not be
considered as resonances in the sense given above.

The formalism leading to the expression for the three-body $S$-matrix
is applicable for a very different system, i.e. three particles of
finite mass with only one non-vanishing two-body interaction. If we
choose the Jacobi coordinate system where $\bm{x}$ is related to the
non-zero two-body interaction the wave function is again given by
Eq.(\ref{eq10}) with a Bessel function instead of $u_{\ell_y}/(yk_y)$.
All derivations remain unchanged and we arrive at the $S$-matrix in
Eq.(\ref{eq17}). The difference is that now only the two-body
$S$-matrix related to the $\bm{x}$-degree of freedom has
poles. Therefore the three-body system has no $S$-matrix poles.

This system effectively also consists of two independent two-body
systems, but now reached in a completely different limit with one
non-zero interaction and arbitrary masses.  The result then illustrates
other aspects of the continuum properties for three-body systems. In
both cases the conclusion is that one subsystem alone cannot produce
three-body resonances or virtual states.  At least two different
subsystems must collaborate and simultaneously contribute with
coinciding poles.  It then seems inevitable that, if all the three
two-body subsystems only have poles, then fully realistic
three-body systems must also have only poles.

\subsection{Particles with spin}
\label{sec4}

The results in the previous subsection remain valid for particles of
non-zero spin if the two-body interactions are spin-independent, because
then the spin part of the wave function is decoupled from the
coordinate part. In fact, even in the case of identical particles the
spin part can always be used to establish the proper (anti)symmetry of
the wave function without changing the coordinate part.  The only
exception is two identical bosons with zero spin where the spin part
of the wave function always is symmetric under exchange of the two
bosons. Therefore odd values of the relative orbital angular momentum
between the bosons are strictly forbidden.  This fact leads to
important differences compared to non-identical particles
\cite{gar04}.

When the two-body interactions are spin-dependent, the separability of
the three-body Hamiltonian for the system shown in Fig.\ref{fig1} can
disappear. We denote the spins of particles $i$, $j$, and $k$ by
$s_i$, $s_j$, and $s_k$, respectively.  Let us for simplicity assume
that the particle-core interactions of the system in Fig.\ref{fig1}
contain a central plus a spin-spin term, i.e.,
$V_{ki}=V_c^{(ki)}+V_{ss}^{(ki)} \bm{j}_i\cdot\bm{s}_k$, where
$\bm{j}_i=\bm{\ell}_{ki}+\bm{s}_i$ and $\bm{\ell}_{ki}$ is the
relative orbital angular momentum between the core and particle $i$
(and similarly for the $V_{kj}$ interaction).  As shown in
\cite{gar03}, this type of spin-spin operator is especially
convenient, since it guarantees conservation of the quantum number
$j_i$ in agreement with the intrinsic motion of particles in the core
possibly being identical to the particles of the three-body system.

The two-body Hamiltonian $H_{ki}$ is then diagonal in the basis
$\left\{ |s_k, (s_i, \ell_{ki})j_i; j_{ki} \rangle \right\}$
where $j_{ki}$ is the total two-body angular momentum after coupling
of $j_i$ to the spin of the core $s_k$. We denote the corresponding
two-body eigenvalues by $E_{ki}^{(\ell_{ki},j_i,j_{ki})}$. The natural 
basis to describe the three-body system is then
$\left\{ |\left[s_k, (s_i, \ell_{ki})j_i\right] j_{ki}, (\ell_{kj},s_j)j_j; J 
\rangle \right\}$, where $J$ is the total three-body angular momentum.
The two-body Hamiltonian $H_{ki}$ with eigenvalues
$E_{ki}^{(\ell_{ki},j_i,j_{ki})}$ is diagonal in this three-body
basis. This is in contrast to the remaining two-body Hamiltonian
$H_{kj}$.

We can calculate the matrix elements of the three-body Hamiltonian in
this basis. With the expression
\begin{eqnarray}
 & & \hspace*{-0.7cm}
|\left[s_k, (s_i, \ell_{ki})j_i\right] j_{ki}, (\ell_{kj},s_j)j_j; J \rangle 
   =  \nonumber \\ & & 
 \sum_{j_{kj}} (-1)^{j_{ki}+j_{kj}+j_i+j_j}
       \sqrt{2 j_{ki}+1} \sqrt{2 j_{kj}+1} \label{eq19} \\ & & 
\times \left\{
    \begin{array}{ccc}
       j_j & s_k & j_{kj} \\
       j_i &  J  & j_{ki}
    \end{array}
    \right\}
|\left[s_k, (\ell_{kj},s_j)j_j\right] j_{kj}, (s_i,\ell_{ki})j_i; J \rangle  
\nonumber
\end{eqnarray}
we obtain by recoupling that
\begin{eqnarray}
& & \hspace*{-0.4cm}
\langle q j_{ki}|H|q^\prime j^\prime_{ki} \rangle =
 \delta_{q q^\prime} \delta_{j_{ki}j^\prime_{ki}} 
                    E_{ki}^{(\ell_{ki},j_i,j_{ki})} \nonumber \\ & &
\hspace*{-4mm}+\delta_{q q^\prime} (-1)^{j_{ki}-j^\prime_{ki}}
\sqrt{(2 j_{ki}+1)(2 j^\prime_{ki}+1)} \label{eq20} \\
&&  \hspace*{-4mm}
\times \sum_{j_{kj}} (2 j_{kj}+1) E_{kj}^{(\ell_{kj},j_j,j_{kj})}
\left\{
    \begin{array}{ccc}
       j_j & s_k & j_{kj} \\
       j_i &  J  & j_{ki}
    \end{array}
    \right\}
\left\{
    \begin{array}{ccc}
       j_j & s_k & j_{kj} \\
       j_i &  J  & j^\prime_{ki}
    \end{array}
    \right\}
\nonumber
\end{eqnarray}
where $q =\{\ell_{ki},\ell_{kj},j_i,j_j\}$ Therefore the three-body
Hamiltonian is diagonal in blocks defined by the quantum numbers $q$.

As an example, we show the $2 \times 2$ block corresponding to a core
and two light particles all three with spin 1/2. We further assume
$j_i=j_j=1/2$, and $J=1/2$ confining $j_{ki}$ and $j_{kj}$ to the
values 0 and 1.  We note that Eq.(\ref{eq20}) depends on the two-body
relative orbital angular momenta, $\ell_{ki}$ and $\ell_{kj}$, only
through the two-body energies.  Then the corresponding block is given
by:
\begin{widetext}
\begin{equation}
(H)=
\left(
    \begin{array}{cc}
 E^{(\ell_{ki},\frac{1}{2},j_{ki}=0)}_{ki}+\frac{1}{4}
   E^{(\ell_{kj},\frac{1}{2},j_{kj}=0)}_{kj}+\frac{3}{4} E^{(\ell_{kj},\frac{1}{2},j_{kj}=1)}_{kj}&
     \frac{\sqrt{3}}{4} 
(E^{(\ell_{kj},\frac{1}{2},j_{kj}=1)}_{kj}-E^{(\ell_{kj},\frac{1}{2},j_{kj}=0)}_{kj})    \\
     \frac{\sqrt{3}}{4} 
(E^{(\ell_{kj},\frac{1}{2},j_{kj}=1)}_{kj}-E^{(\ell_{kj},\frac{1}{2},j_{kj}=0)}_{kj}) &     
E^{(\ell_{ki},\frac{1}{2},j_{ki}=1)}_{ki}+\frac{3}{4}
   E^{(\ell_{kj},\frac{1}{2},j_{kj}=0)}_{kj}+\frac{1}{4}E^{(\ell_{kj},\frac{1}{2},j_{kj}=1)}_{kj}
    \end{array}
    \right)
\label{eq21}
\end{equation}
\end{widetext}

This illustrates that even for the schematic case in Fig.\ref{fig1},
where particles $i$ and $j$ do not interact with each other, the
three-body Hamiltonian is not separable anymore. The three-body
eigenvalues are then not given by the sum of the two-body
eigenvalues. When particles $i$ and $j$ are identical and
$\ell_{ki}$=$\ell_{kj}$=$\ell$, then
$E_{ki}^{(\ell,\frac{1}{2},j_{ki}=0)}$=
$E_{kj}^{(\ell,\frac{1}{2},j_{kj}=0)}$=$E^{(0)}$ and
$E_{ki}^{(\ell,\frac{1}{2},j_{ki}=1)}$=
$E_{kj}^{(\ell,\frac{1}{2},j_{kj}=1)}$=$E^{(1)}$, and the eigenvalues
of the matrix (\ref{eq21}) are
$\frac{1}{2}E^{(0)}$+$\frac{3}{2}E^{(1)}$ and
$\frac{3}{2}E^{(0)}$+$\frac{1}{2}E^{(1)}$. Only the first of these
eigenvalues corresponds to an antisymmetric eigenfunction under
exchange of particles $i$ and $j$.  It is precisely twice the average
energy of the two two-body energies.  Due to the Pauli principle, the
two light fermions must occupy both the two possible two-body states
with angular momentum $\frac{1}{2}$.  Then the relevant energies are
not the individual energies of the two two-body states, but their
average value.

When none of the two-body interactions contains the spin-spin term we
have that
\begin{eqnarray}
E_{ki}^{(\ell_{ki},\frac{1}{2},j_{ki}=0)}&=&
E_{ki}^{(\ell_{ki},\frac{1}{2},j_{ki}=1)} \equiv
E_{ki}^{(\ell_{ki},\frac{1}{2})} \; , \label{eq22a} \\ \label{eq22b}
E_{kj}^{(\ell_{kj},\frac{1}{2},j_{kj}=0)}&=&
E_{kj}^{(\ell_{kj},\frac{1}{2},j_{kj}=1)} \equiv
E_{kj}^{(\ell_{kj},\frac{1}{2})} \; ,
\end{eqnarray}
and the Hamiltonian (\ref{eq21}) is diagonal, with identical diagonal
terms given by $E_{ki}^{(\ell_{ki},\frac{1}{2})}$+
$E_{kj}^{(\ell_{kj},\frac{1}{2})}$, which is the result obtained in
the previous section for spin-zero systems.  When only one of the
two-body interactions has a non-zero spin-spin term (e.g., $V_{ki}$,
Eq.(\ref{eq22a}) is invalid and Eq.(\ref{eq22b}) is valid) the
Hamiltonian is diagonal in the basis $\left\{ |\left[s_k, (s_i,
\ell_{ki})j_i\right] j_{ki}, (\ell_{kj},s_j)j_j; J\rangle \right\} $
with the energies
$E_{ki}^{(\ell_{ki},\frac{1}{2},j_{ki}=0)}$+$E_{kj}^{(\ell_{kj},\frac{1}{2})}$
and
$E_{ki}^{(\ell_{ki},\frac{1}{2},j_{ki}=1)}$+$E_{kj}^{(\ell_{kj},\frac{1}{2})}$.
There are then two possible three-body states, that again are given by
the sum of the two-body energies.

\section{General properties}

\begin{figure}
\vspace*{-19cm}
\begin{center}
\epsfig{file=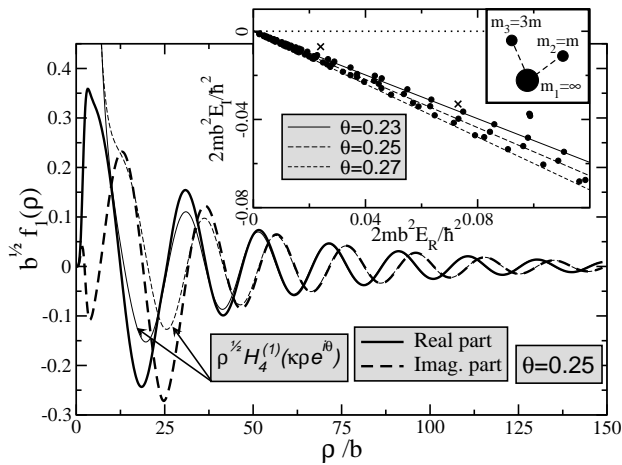,scale=1.0,angle=0}
\end{center}
\caption[]{Three-body resonance energy and radial wave function
for $V_{23}$=0 and two gaussian $p$-wave interactions $V_{12}$ and
$V_{13}$ with ranges $b_{12}$=1.5$b$ and $b_{13}$=$b$.
The strengths produce resonances at $2 m b^2E_{12}/\hbar^2$=0.073$-i$0.033,
$2 m b^2E_{13}/\hbar^2$=0.024$-i$ 0.007.  The masses are $m_1$=$\infty$,
$m_2$=$m$, $m_3$=3$m$.
Small graph: The solid,
long-dashed and short-dashed lines connect the computed
continuum spectra for rotation angles $\theta$=0.23, 0.25, 0.27,
respectively.  The circle above these lines is a three-body
resonance independent of $\theta$.  Big graph: The thick lines are the
real (solid) and imaginary (dashed) parts of the lowest hyperradial
adiabatic
wave function of the three-body resonance when $\theta$=0.25.
The corresponding thin (solid, dashed) lines show the asymptotic
behaviour.}
\label{fig2}
\end{figure}

The results obtained in the previous section for the system in
Fig.\ref{fig1} can be taken as a test for the numerical method used to
compute three-body states. Taking then the schematic system as
starting point we analyze the main properties of the three-body
states, and how they evolve in the energy plane when different
ingredients are added to the calculations.

\subsection{Bound states and resonances}

We maintain the schematic model where $V_{23}$=0 and
$m_1$=$\infty$. We first assume that both $V_{12}$ and $V_{13}$ only
act in relative $p$-waves producing resonances of complex energies
$E_{12}$ and $E_{13}$, respectively.  The three-body computation then
leads to the results shown in Fig.~\ref{fig2} for a specific set of
parameters. In the small panel, together with the discretized complex
rotated three-body continuum states (plotted for three different
scaling angles), there are two additional complex rotated branch cuts
starting at each of the two-body resonance energies (crosses in the
figure).  These branch cuts correspond to two of the particles in a
two-body resonant state and the third particle in the continuum. To
keep the figure cleaner these cuts have not been plotted. Taking $b$
as an arbitrary length unit, the three-body system has a resonance at
the energy $2 m b^2E/\hbar^2$=0.097$-i$0.040, clearly distinguishable
from the continuum background and independent of the rotation
angle. The computed energy is precisely at $E$=$E_{12}$+$E_{13}$, as
found in section III. This numerical result is then proving the
efficiency of the numerical method. To illustrate that this indeed is
a three-body resonance we also show the lowest radial wave function
computed for a given rotation angle $\theta$. Both the real and
imaginary parts vanish asymptotically following the corresponding
Hankel function $H_{K+2}^{(1)}$ as required for a pole of the
$S$-matrix as seen from Eq.(\ref{eq6}).

\begin{figure}
\vspace*{-4.1cm}
\begin{center}
\epsfig{file=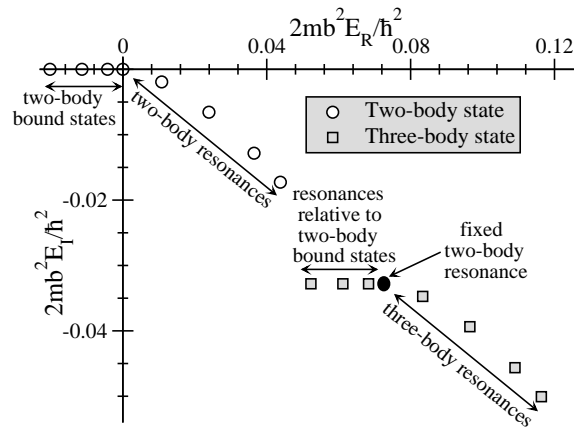,scale=1.0,angle=0}
\end{center}
\vspace*{-13.1cm}
\caption[]{Real ($E_R$) and imaginary ($E_I$) energy ($E=E_R + iE_I$)
of the three-body states (squares) for a system with parameters as in
Fig.~\ref{fig2}. The resulting energy is the sum of a fixed two-body
resonance energy (black circle) and the energy of a varying two-body
state (open circles). When this two-body state is
bound the three-body state is the two-body resonance relative to the
bound state.}
\label{fig3}
\end{figure}

When one of the two-body systems is bound and the other has a
resonance the three-body state with energy equal to the sum of the
two-body energies is simply a two-body resonance of one particle
relative to the two-body bound state.  This is illustrated in
Fig.~\ref{fig3}, where one two-body resonance remains unchanged while
the other two-body state is varied from resonance to bound state.
When one two-body system is bound, and thus appearing on the negative
real energy axis (open circles), the three-body energies (squares)
appear at the same distances to the left of the fixed two-body
resonance (black circle).

If the attractive interaction binding the two-body system decreases,
at some point the two-body state enters through the origin into the
fourth quadrant of the energy plane and becomes a two-body resonance.
In parallel, the three-body energy (squares) approaches the two-body
resonance energy (black circle) and continues through the two-body
resonance in southeastern direction all the time following the
addition rule.

\subsection{Virtual states}

The complex scaling method can be viewed as an analytic continuation
into complex coordinates, such that resonances are ``pulled out'' of
the continuum.  Unfortunately the method fails when the pole
corresponds to a virtual state, due to the necessary large rotation
angle.  Virtual states remain in the unphysical Riemann sheet, and
numerical investigation of their effects is more difficult.

\begin{figure}
\vspace*{-18.5cm}
\begin{center}
\epsfig{file=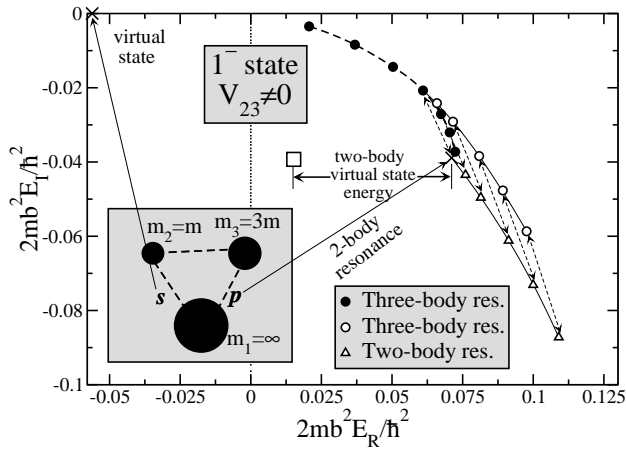,scale=1.0,angle=0}
\end{center}
\caption[]{Energies of the 1$^-$ three-body states for a system with the same
masses as in Fig.~\ref{fig2}. The $p$-wave and $s$-wave interactions,
$V_{13}$ and $V_{12}$, correspond to a resonance and a virtual state
indicated by the crosses. The thick open square is the sum of the two
two-body energies. Inclusion of a varying attractive $s$-wave interaction
$V_{23}$ leads to the three-body resonance energies given by the solid circles.
For a fixed value of $V_{23}$, when the two-body $p$-resonance moves as shown
by the open triangles, then the three-body resonances do it as given by the
open circles.}
\label{fig4}
\end{figure}

Let now the interactions $V_{12}$ and $V_{13}$ correspond to a virtual
$s$-state and a $p$-resonance of energies $E_{12}$ and $E_{13}$,
respectively ($V_{23}$=0, $m_1$=$\infty$). Then, a 1$^-$ three-body
$S$-matrix pole should be present at $E$=$E_{12}$+$E_{13}$.  To test
this numerically we include an attractive $s$-wave interaction
$V_{23}$ which, combined with the other interactions, is sufficiently
strong to reveal the existence of a 1$^-$ three-body resonance. We
show in Fig.~\ref{fig4} how the complex energy of this resonance moves
as the strength of $V_{23}$ is varied.  The strongest attraction (the
point closest to the origin) corresponds to a virtual state in the 2-3
subsystem at an energy of about $-0.004$ in Fig.\ref{fig4}. A very small
additional attraction would bind the three-body system, which would be
Borromean if the three-body system becomes bound before the 2-3
subsystem.

When the strength of $V_{23}$ decreases the three-body resonance moves
(closed circles) towards the energy $E_{13}$ of the $p$-resonance. If
we could track the three-body resonance for even weaker $V_{23}$ we
would for some non-zero value find that it precisely coincides with
$E_{13}$ (cross), reaching then the discontinuity cut of the Riemann
sheet. For weaker $V_{23}$ the three-body pole could move 
through this cut into any other Riemann sheet, but it is tempting
to suggest that these states could be interpreted as two-body virtual 
$s$-states of
particle $2$ relative to the resonant state of particles $1$ and
$3$. Thus such a three-body state is not a three-body resonance but a
virtual state on the unphysical Riemann sheet, although it is not
possible to see numerically how this energy (square) is reached when
$V_{23}$=0.

This can be better understood if the $s$-state is bound as illustrated
in Fig.~\ref{fig3}. When the $s$-wave attraction is reduced until the
bound state energy is zero the three-body state moves in parallel
towards the resonance energy. A continued decrease of the attraction
turns the two-body bound state into a virtual $s$-state, and the
two-body pole moves continuously from the physical to the unphysical
Riemann sheet.  In parallel, the three-body pole moves continuously
through the $p$-resonance onto another Riemann sheet.

The connection between two and three-body resonances are very intimate
even when all three masses and interactions are non-zero.  This is
emphasized in Fig.\ref{fig4} where $V_{13}$ is varied for fixed
$V_{23}$.  Then the 1$^-$ three-body resonances (open circles) follow
precisely the motion of the two-body resonances (open triangles).  The
double arrows connect each two-body resonance with its corresponding
three-body state. A variation of the two-body complex energy produces
a similar energy change in the three-body state.

For completeness we also notice that virtual states in both subsystems
lead to virtual three-body states with an energy following the sum
rule and sitting in the negative energy axis of the unphysical Riemann
sheet.

\subsection{Finite mass and angular momentum coupling}

In a less schematic model also $m_1$ has to be finite. Still often the
three-body structure is mostly influenced by two two-body subsystems
each dominated by one partial wave. Thus we relax the condition
$m_1$=$\infty$ but maintain the reduced masses $\mu_{12}$ and
$\mu_{13}$ by adjusting $m_2$ and $m_3$. The two-body properties then
remain unchanged for the same interactions $V_{12}$ and $V_{13}$.  The
finite masses destroy the separability into two independent
subsystems, and the eigenvalues of the three-body Hamiltonian are not
any more given by the sum of the two-body eigenvalues. Also, a given
combination of partial waves may couple to several total angular
momenta with different energies.

\begin{figure}
\vspace*{2cm}
\begin{center}
\epsfig{file=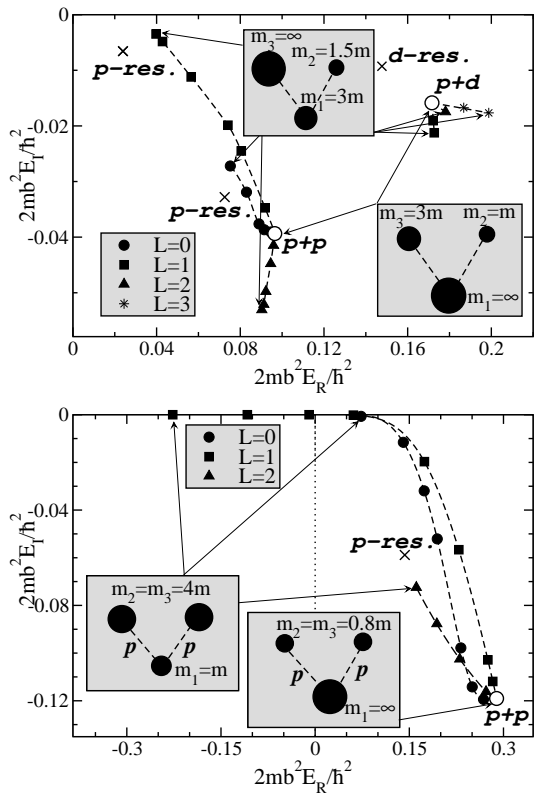,scale=0.55,angle=0}
\end{center}
\caption[]{Upper part: Three-body resonances for a system with two-body
reduced masses and resonance parameters as in Fig.~\ref{fig2} ($p$+$p$
case). In the $p$+$d$ case the highest $p$-resonance has been substituted
by a $d$-resonance with $2 m b^2E_{12}/\hbar^2$=0.145-$i$ 0.009.
Lower part: Three-body resonances for a system with
$\mu_{12}$=$\mu_{13}$=0.8$m$ and $p$-resonances at
$E_{12}$=$E_{13}$ with $2 m b^2E_{12}/\hbar^2$=0.143-$i$ 0.059.
The fixed two-body
resonances are indicated by the crosses and the three-body resonances
by a big open circle ($m_1$=$\infty$), and solid circles ($L$=0),
squares ($L$=1), triangles ($L$=2), and stars ($L$=3) for finite
$m_1$ values. }
\label{fig5}
\end{figure}

When two partial $p$-waves both contribute, the total angular momentum
and parity must be $L^{\pi}$=0$^+$, 1$^+$, 2$^+$, and combining a
$p$-wave and a $d$-wave we get $L^{\pi}$=1$^-$, 2$^-$, 3$^-$. The
corresponding three-fold degeneracy is broken for finite $m_1$ as
illustrated in the upper part of Fig.~\ref{fig5}. In general, for fixed 
structure of the system (fixed reduced masses, interactions, relative 
distances between particles...), the three-body relative kinetic energy is 
smaller for finite mass of the core than for infinite mass, as seen by 
comparing $p_{ik}^2/2\mu_{ik}+p_{j,ik}^2/2\mu_{i,jk}$ in both cases.  Thus the 
three-body resonances often tend to move towards the origin.

However, simultaneous conservation of angular momentum and
center-of-mass might require a change of structure resulting in less
total interaction energy than from the two resonances. Thus, the
three-body resonance can move in all directions (Fig.~\ref{fig5}). 
As in the previous calculations, only the components corresponding to
two-body resonances are included. The
finite mass effects are relatively small in all cases except for two
$p$-waves coupled to $L^{\pi}$=1$^+$. This can be understood by
expressing the Faddeev component related to the 1-3 subsystem in the
Jacobi coordinates of subsystem 1-2.  For $L$=1 a $p$-wave ``rotates''
fully into a $p$-wave.  Then the $p$-wave attraction is fully
exploited and the resulting three-body state has the lowest possible
energy.  When $L$ and the two two-body angular momenta are equal there
is a similar tendency, decreasing with $L$, to maximize the total
attraction.  Other components obtained by rotation are less important
and uneven two-body angular momenta cannot exploit the attraction as
efficiently, producing a much smaller change with $m_1$.  In practice
detailed quantum mechanical computations are needed for each case to
determine the size of the effects.

It is remarkable that the effects of the coupling and the
center-of-mass motion can lower the energy sufficiently to produce a
1$^+$ three-body bound state (lower part of Fig.~\ref{fig5}). This is
a Borromean system obtained by only two two-body interactions.  When
one resonance is replaced by a virtual state we have not numerically
been able to find any three-body resonance produced entirely by finite
mass effects and without change of interactions.

\section{Numerical illustrations}

The origin of the resonances and bound states can be illustrated by
different examples connecting the bare schematic model with well known
realistic nuclear structures. We first study two nucleons outside the
spin-zero $\alpha$-particle core, then we present detailed realistic
computations for the same $A=6$ systems, and finally we extend to the
non-zero core-spin of the Borromean nucleus $^{17}$Ne.

\subsection{Core with zero spin}

\begin{figure}
\begin{center}
\epsfig{file=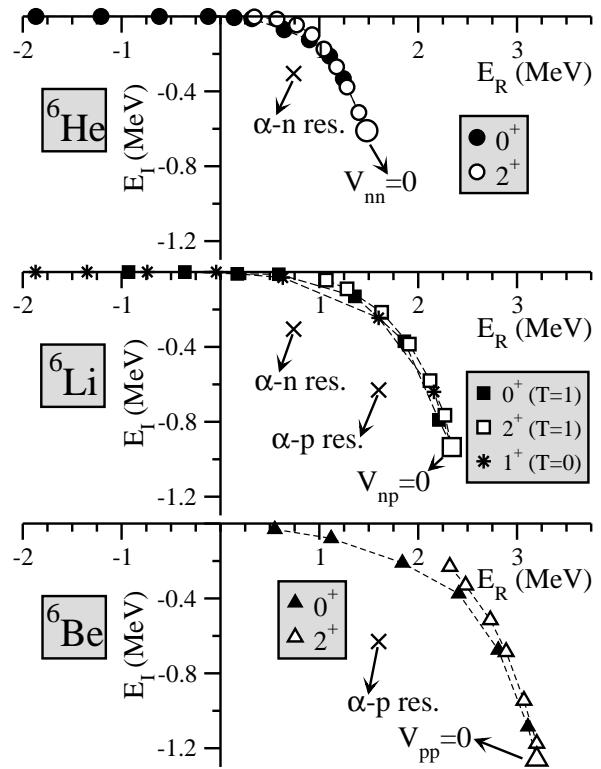,scale=0.5,angle=-90}
\end{center} 
\caption[]{Evolution of the three-body resonances for $^6$He (circles), 
$^6$Li (squares), and $^6$Be (triangles), when the nucleon-nucleon 
interaction is progressively introduced. The big open symbols correspond
to zero nucleon-nucleon interaction. The 0$^+$ and 2$^+$ resonances are
shown by the closed and open symbols respectively. The stars in 
the $^6$Li case are the 1$^+$ resonance corresponding to the $T=0$ channel
in the neutron-proton interaction. The crosses show the alpha-nucleon
resonance energies.}
\label{fig6}
\end{figure}

Let us consider a system made by an infinitely heavy core and two
non-interacting spin 1/2 particles. The core is assumed to be charged
(with twice the proton charge) and the particle-core reduced mass is
0.8$m$, where $m$ is the nucleon mass. This system is very similar to
an $\alpha$-particle and two nucleons. The particle-core reduced mass is
the same, and the center-of-mass effects are very small and not
relevant for our present purpose \cite{gar04}.

Let us also consider a simple particle-core interaction given by a
central short-range $p$-wave interaction producing a particle-core
(uncharged) resonance at an energy of 0.74 MeV with a width of 0.60
MeV.  When the light particle is assumed to have the proton charge,
the same short-range interaction produces a core-particle resonance
with energy and width of 1.60 MeV and 1.26 MeV, respectively. These
values are consistent with the lowest $p$-resonances in $^5$He and
$^5$Li, respectively \cite{ajz88}.

In Fig.\ref{fig6}, we show the results for one heavy and two light
particles resembling the $^6$He, $^6$Li, and $^6$Be systems.  The
upper, middle and lower parts correspond to both uncharged, one
neutral and one with the proton charge, and both with the proton
charge, respectively.  When the nucleon-nucleon interaction is equal
to zero, the three-body system must have a resonance at an energy
equal to the sum of the two two-body resonance energies.  This is
again confirmed by calculations using the complex rotated adiabatic
expansion method.  We obtain the three-body resonances indicated by
the big open circle, big open square, and big open triangle in the
upper, central, and lower parts of Fig.\ref{fig6}. They match
precisely with the sum of the two-body energies each indicated by a
cross.

The known spectrum for the schematic three-body system, determined
completely by the internal two-body states, can be used as starting
point to trace the connection with the ``realistic" three-body
system. This must be done by including the effects produced by i) an
additional interaction between the two light particles, ii) quantum
numbers conservation (like the angular momentum), and iii)
center-of-mass effects originating from the finite mass of the core.

Effects i) and ii) are closely connected. Once the interaction between
the two light particles is included the three-body wave function,
initially independent of the total angular momentum (Eq.(\ref{eq12})),
is now depending on $L$. This can be seen in the upper part of
Fig.\ref{fig6}, where we include the neutron-neutron interaction
multiplied by a global factor varying from 0 to 1.  Then the 0$^+$ and 2$^+$
states evolve as shown by the solid and open circles,
respectively. When the full neutron-neutron interaction is included
(last point on each curve), the system similar to $^6$He has a bound
Borromean 0$^+$ state (with a binding energy close to $-1.9$ MeV) and
a very narrow 2$^+$ three-body resonance with energy 0.34 MeV.

For the system similar to $^6$Li shown in the middle part of the
figure one of the light particles has the charge of the proton.  The
neutron-proton interaction is again continuously switched on from zero
to full strength.  The 0$^+$ and 2$^+$ states ($T$=1), analogous to
those in the upper part, move as given by the closed and open squares,
respectively. The final 0$^+$ state is still below the three-body
threshold, with a binding energy of $-1.0$ MeV. The final $2^+$
resonance has an energy of 1.1 MeV and a width of 0.1 MeV. The shift
in energy compared to the states in the upper part is produced by the
Coulomb repulsion between the core and the charged particle. Now the
allowed $T$=0 coupling gives rise to additional three-body states,
e.g.  the 1$^+$ state shown by the stars.  When the
three-body 1$^+$ state becomes bound the system is still
Borromean. However, for some threshold strength the two light
particles become bound, and the system is not Borromean anymore.  When
the full interaction is included the two light particles then form a
deuteron nucleus, and the three-body 1$^+$ state becomes the ground
state of the system, very similar to $^6$Li, with a binding of about
$-5.5$ MeV (outside the scale of the figure).

In the lower part of the figure we show results for two light
particles each with the proton charge.  As in the upper part only the
$T=1$ states are allowed, and since all the three two-body subsystems
feel the Coulomb repulsion, the energies of the $0^+$ states (closed
triangles) and the $2^+$ states (open triangles) are clearly larger
than in the previous cases. In fact, when the full proton-proton
interaction is included the $0^+$ ground state is unbound, with
resonance energy and width of 0.5 MeV and 0.2 MeV, respectively. For
the 2$^+$ resonance the energy and width are 2.3 MeV and 0.5 MeV.

\begin{table}
\begin{tabular}{|cc|cccc|}
\hline
       &        &   Scheme    & Real.(no 3-b)  &  Realistic & Exper.\\ \hline
$^6$He  & 0$^+$ & $-1.87$     & $-0.04$ &  $-0.96$ &  $-0.97\pm0.04$ \\ 
        & 2$^+$ & (0.34,0.01) & (1.36,1.12)  & (0.87,0.11) & (0.83,0.11) \\ 
                                                                       \hline
        & 0$^+$ & $-0.93$     & (0.75,0.15) & $-0.14$ & $-0.14$  \\ 
$^6$Li  & 1$^+$ & $-5.60$     & $-3.05$ & $-3.73$  & $-3.70$  \\ 
        & 2$^+$ & (1.07,0.09) & (1.92,0.87) & (1.66,0.50) & (1.67,0.54) \\ 
                                                                       \hline
$^6$Be  & 0$^+$ & (0.55,0.16) & (2.05,0.58) & (1.37,0.11) & (1.37,0.09)  \\ 
        & 2$^+$ & (2.32,0.46) & (3.10,1.89) & (3.02,1.65) & (3.04,1.16)  \\ 
                                                                       \hline
\end{tabular}
\caption{Computed and experimental energies (in MeV) for $^6$He, $^6$Li, 
and $^6$Be. The energies within brackets are $(E_R,\Gamma_R)$, that give
the resonance energy and its width. The computations correspond to the 
schematic system described in the text (second column), realistic calculations
without inclusion of an effective three-body force (third column), and the same
calculation when a three-body force is used (fourth column). The experimental 
values (last column) are taken from \cite{ajz88}. When not specified, the error
bars are smaller than the last digit.}
\label{tab1}
\end{table}

The effects produced by the finite mass of the core (center-of-mass
effect), do not change the previous energies significantly (typically
no more than 100 keV \cite{gar04}). The spectra obtained for the
``simplified" $^6$He, $^6$Li, and $^6$Be nuclei (only a simple
$p$-wave particle-core interaction has been used), indicate where to
find the true states. This is seen in table~\ref{tab1}, where the
second column gives the computed energies.  A comparison with the
experimental results (last column) reveals that in all the three
nuclei the main features of the spectrum are well reproduced.
Therefore, starting from the schematic case where the three-body
energy is given by the sum of the two-body energies and taking into
account the main characteristics of the remaining two-body
interaction, it is possible to estimate where the three-body states
must be placed.

As seen in the table, the computed energies are always lower than the
experimental ones. This is because the nucleon-core interaction is
without any spin-orbit term, and consequently the two possible
$p$-resonances ($p_{1/2}$ and $p_{3/2}$) appear at the same energy.
This is overbinding the system, since the $p_{1/2}$-resonance is known
to be a few MeV higher \cite{ajz88}, and a realistic calculation must
include a repulsive term pushing up this resonance to the correct
value.  On top of this, other than $p$-waves should be included in the
calculation.  For $s$-waves the effect of the Pauli principle must be
accounted for, since the $s_{1/2}$-shell is fully occupied in the
$\alpha$-core.  This effectively amounts to a highly repulsive
$s$-wave potential at short distances, also pushing the computed
energies towards higher values.  Furthermore, it is well known that
three-body calculations using only realistic two-body interactions
typically underbind the system.  This can be cured by inclusion of an
additional potential taking into account possible three-body effects
to reproduce the experimental values.

Realistic detailed calculations using the (complex scaled) hyperspheric
adiabatic expansion method concerning $^6$He can be found in 
\cite{gar99,fed03}. For completeness we show in the next section the 
results obtained for $^6$Li and $^6$Be when similar calculations are
performed.

\subsection{Realistic calculations for $^6$Li and $^6$Be}

In \cite{gar99} the use of phase equivalent potentials is suggested as
an efficient method to take into account the Pauli principle in
three-body calculations. This method, together with the hyperspheric
adiabatic expansion method, is used to compute the ground state of
$^{11}$Li and $^6$He.  Thus, for $^6$He the alpha-neutron $s$-wave
interaction is able to bind the neutron into a Pauli forbidden
state. To exclude this state we use a phase equivalent potential with
exactly the same phase shifts for all energies, but with one less
bound state.  The $p$-wave interaction contains a central and a
spin-orbit term, such that the $p_{3/2}$ and $p_{1/2}$ resonances in
$^5$He are placed at the experimental energy values \cite{ajz88}. The
neutron-neutron interaction can also be found in \cite{gar99}. 

The calculation using only the two-body interactions underbinds the
three-body nucleus (third column in table~\ref{tab1}), and an
additional effective three-body force is needed to recover the
experimental value (fourth column).  In \cite{fed03} the same method
supplemented by complex scaling is used to compute the 2$^+$ resonance
in $^6$He. In the present work the calculation is slightly simplified
by using a repulsive $s$-wave alpha-neutron interaction to take the
Pauli principle into account. When the same phase equivalent potential
as for the ground state is used, the computed 2$^+$ state in $^6$He is
also given in the third and fourth columns of table~\ref{tab1}. As
before, the calculation without a three-body force is underbinding the
system by about 0.5 MeV, and an effective three-body force is again
needed. In all the calculations the three-body force is assumed to be
gaussian with a range of 3 fm and the strength is adjusted to fit the
experimental value. For bound states it is obviously possible to find
a strength fitting precisely the experimental value. For resonances,
this single parameter is fitting simultaneously the energy and the
width of the resonance.

For $^6$Li and $^6$Be we perform exactly the same calculations but
adding the corresponding Coulomb interaction for the two cases. The
additional proton-core repulsion in $^6$Li is pushing up the energies
of the $0^+$ and 2$^+$ states compared to the energies obtained for
$^6$He, and even more for $^6$Be, where all the three two-body
subsystems feel the Coulomb repulsion. The computed energies are again
given in the third column of table~\ref{tab1} when only the two-body
interactions are used. As before, an effective three-body interaction
is needed to fit the experimental energies (fourth column). The $0^+$
state in $^6$Li is still below the two-nucleon separation energy
threshold, although when the three-body interaction is suppressed the
computed state appears to be unbound. Again, when the strength of the
three-body force is used to match the experimental resonance energies,
the widths are also well reproduced, except for the $2^+$ state in
$^6$Be, where the width is significantly larger than the experimental value.

In addition to the $0^+$ and 2$^+$ $T=1$ states, in $^6$Li 
it is also possible to have states with $T=0$.  These three-body states
correspond to structures where the neutron and the proton form a
deuteron ($d$) nucleus, suggesting a description of the nucleus as a
$d$+$\alpha$ two-body system. Therefore, we have performed a
three-body calculation, identical to the ones described above, but
where the nucleon-nucleon interaction has been substituted by the
$T=0$ interaction used in \cite{cob97}.  This potential reproduces the
experimental deuteron binding energy, root mean square radius,
electric quadrupole moment, and provides a $d$-wave content of
5.6\%. The correspondig computed energies for the $1^+$ (ground) state
are also given in table~\ref{tab1}.

The three-body Thomas-Ehrman shifts \cite{gar03a} of these isobaric
analog $0^+$ and $2^+$ states are then obtained from table~\ref{tab1},
i.e. in MeV (1.40,1.17,1.05) and (1.83,1.80,1.65) for the realistic
cases without and with three-body potential, respectively. The latter
results coincide by definition with the experimental values.  The
decreasing tendency in the computed differences are due to the smaller
energy and the larger radii which in turn leads to smaller effects of
the Coulomb interaction. The underbinding is responsible for this
discrepancy with measurements.

\subsection{Core with non-zero spin}

When the three particles have non-zero spins and spin-dependent two-body
interactions, as in any realistic nuclear potential, then the
three-body Hamiltonian is not even separable in the schematic model
with an infinitely heavy core and two mutually non-interacting
particles, see section \ref{sec4}.  The Hamiltonian matrix is then
organized in blocks with the matrix elements given by
Eq.(\ref{eq20}).

An example is for $^{17}$Ne, that is well described as a three-body
system made by an $^{15}$O core and two protons \cite{gar04b}. The
core of the system has negative parity and spin 1/2. The two-body
subsystem, $^{16}$F ($^{15}$O+$p$), has four low-lying resonances, two
of them ($0^-$ and $1^-$) arise from the coupling of a relative
$s_{1/2}$-wave and the spin of the core, and other two (2$^-$ and
$3^-$) result from the coupling between a $d_{5/2}$-wave and the spin
of the core.  The experimental energies and widths ($E_R$,$\Gamma_R$)
of these two-body resonances are (0.535,0.040$\pm$0.020) MeV,
(0.728$\pm$0.006,$\langle$0.040) MeV, (0.959$\pm$0.005,
0.040$\pm$0.030) MeV, and (1.256$\pm$0.004, $\langle$0.004) MeV,
respectively \cite{ajz86}.

In \cite{gar04b} several proton-core interactions, all of them reproducing 
the experimental energies and widths, are given.  These
interactions contain spin-dependent operators, in particular the spin-splitting
operator $\bm{s}_c \cdot \bm{j}_p$, where $\bm{s}_c$ is the spin of the
core and $\bm{j}_p$ results from the coupling between the relative
orbital angular momentum and the spin 1/2 of the proton. In the following
calculations we shall use the $s$ and $d$-wave interactions corresponding to 
a gaussian two-body potential with the parameters given in Table 1 of 
\cite{gar04b}, but where the spin-orbit strength has been modified to push 
away the $d_{3/2}$-resonances in $^{16}$F, but keeping the two
$d_{5/2}$-resonances at the right energy. In this way our slightly
simplified $^{17}$Ne is characterized exclusively by the $0^-$, $1^-$,
$2^-$, and $3^-$ resonances in $^{16}$F, and we can perform for $^{17}$Ne
the same kind of analysis as in the previous cases.

\begin{figure}
\vspace*{-18.5cm}
\begin{center}
\epsfig{file=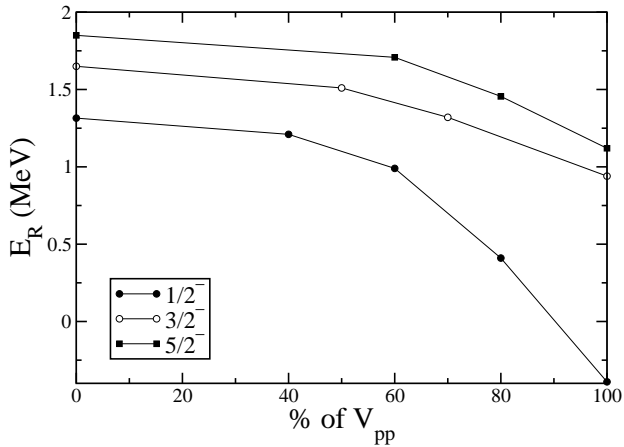,scale=1.0,angle=0}
\end{center}
\caption[]{$^{17}$Ne energies for the 1/2$^-$, 3/2$^-$, and 5/2$^-$ states
as a function of the percentage of the proton-proton interaction included
in the calculation.}
\label{fig7}
\end{figure}

Let us begin by performing calculations for the schematic model
without any proton-proton interaction and for an infinitely heavy
core. The effect of the finite mass of the core is clearly smaller
than for $^6$He, $^6$Li, and $^6$Be, and the core can safely be
assumed to be infinitely heavy.

When both protons are in an $s$-wave relative to the core, only the
$0^-$ and $1^-$ states in $^{16}$F are obtained, and only the components
with relative core-proton orbital angular momentum equal to zero are included
in the calculation. In this case the
total three-body angular momentum can only take the value $J=1/2$
\cite{gar04b}, and the block to be diagonalized is given in
Eq.(\ref{eq21}). From the two possible eigenfunctions only one of them
is antisymmetric under exchange of the two protons, and corresponds to
the eigenvalue $(E^{(0^-)}+3E^{(1^-)})/2$, that is twice the average
energy of the two $s_{1/2}$-resonances. This gives rise to a $1/2^-$
energy of 1.32 MeV, that is recovered when the calculation using the
complex scaled hyperspheric adiabatic expansion method is performed.

When the proton-proton interaction $V_{pp}$ is progressively
introduced, the contribution from the $d$-waves starts to be relevant,
and the components with relative core-proton orbital angular momentum equal
to 2 are also included. The energy of the $1/2^-$ state then changes as shown 
by the solid circles in Fig.\ref{fig7}. Around 90\% of the proton-proton
interaction is already binding the three-body system, becoming then Borromean. 
The full proton-proton interaction gives a binding
of about 0.40 MeV and a $d$-wave content of 37\%. This binding is 0.3
MeV smaller than the one given in table 8 of \cite{gar04b}. This is
due to the fact that in this calculation, in order to clean the structure
of $^{17}$Ne,  we have artificially pushed up the
$d_{3/2}$-resonances by using a large spin-orbit splitting. This
is also reducing the $d$-wave content compared to the result in
\cite{gar04b}.

In ref.\cite{gar04b} we show how the 3/2$^-$ state in $^{17}$Ne is
dominated by the $sd$-interferences between the components. To investigate
this resonances both (0 and 2) core-proton relative orbital angular momenta
are then needed and therefore included in the calculation. For the
schematic $^{17}$Ne model ($V_{pp}=0$), one of the protons can be in
one of the $s$-resonances in $^{16}$F (0$^-$ or 1$^-$), and the other
can be in one of the $d_{5/2}$-resonances ($2^-$ or $3^-$). Assuming
$J=3/2$, the block in Eq.(\ref{eq20}) is already diagonal
with only one non-zero energy equal to $E^{(1^-)}+E^{(2^-)}$.  This
energy of 1.65 MeV is recovered in the numerical calculation, which
also reveals that the only non-vanishing components correspond to one
proton in the $s$-wave $1^-$ resonance and the other in the $d$-wave
$2^-$ resonance.  The evolution of the energy when the proton-proton
interaction is introduced is shown by the white circles in
Fig.\ref{fig7}, reaching a final value of 0.94 MeV. This value is also
around 0.3 MeV less bound than the one given in table 8 of
\cite{gar04b}.

Finally, in the figure we also show the evolution of the lowest
5/2$^-$ resonance in $^{17}$Ne (solid squares) when $V_{pp}$ is
introduced. This state is also dominated by $sd$-interference
components.

\section{Qualitative considerations}

Using the schematic case in which $V_{23}=0$ as starting point, it is
possible to trace the behaviour of the three-body resonances
(Figs. \ref{fig2}, \ref{fig3}, and \ref{fig4}) when the center-of-mass
effects and the $V_{23}$ interaction are included. Taking into account
the main features of the $V_{23}$ interaction, it is then possible to
make crude estimates concerning different realistic systems.

For nuclear systems we take the mass unit $m$ equal to the nucleon
mass and a length unit equal to a typical range for the nuclear
interaction ($b=2$ fm). With this choice, the lowest $p$-resonance in
the neutron-alpha interaction has dimensionless energy $2m b^2 |E_R| /
\hbar^2 \approx 0.15$.  Therefore, in $^6$He ($\alpha$+$n$+$n$),
before including the neutron-neutron interaction, the three-body
system must have a resonance with dimensionless energy of about
0.30. The neutron-neutron interaction has a low lying virtual state at
$2m b^2 |E_R| / \hbar^2 \approx 0.02$, which is enough to bind the
three-body state as evidenced by the bound state in $^6$He at $2m b^2
E_R / \hbar^2 \approx -0.19$.

For $^{11}$Li ($^{9}$Li+$n$+$n$) the $s$-waves are pronounced and the
lowest neutron-$^{9}$Li virtual $s$-state would after spin averaging
appear at about 0.07 which, combined with the neutron-neutron
interaction, is sufficient to bind the three-body system.  Substituting
$^{9}$Li by a $\Lambda$-particle we arrive at an unstable system. With
the $\Lambda$-nucleon $s$-wave scattering length of 2~fm obtained from
hypertriton computations \cite{cob97} the virtual $s$-state would be
at around 1.7, too large to bind the $\Lambda$-neutron-neutron system.

We can continue to a more speculative system obtained by combining the
recently discovered pentaquark resonance with a third particle like a
meson. The pentaquark, $\theta^+$, has a mass of 1540 MeV and decays
mainly into a kaon $K^+$ and a neutron \cite{sch03}.  The energy above
threshold is then around 107~MeV, that corresponds to a dimensionless
energy of about 20. Thus, the neutron-neutron interaction, that was
able to bind $\alpha$+$n$+$n$ and $^{9}$Li+$n$+$n$ respectively with
0.15 and 0.07, but not $\Lambda$+$n$+$n$ with 1.7, cannot bind the
$K^++n+n$ system.  Thus a nucleon-kaon-meson cluster system would also
exploit the intrinsic quark degrees of freedom and not resemble any
three-body structure although additional binding would be picked up.
In any case the narrow width of $\theta^+$ strongly suggests that the
resonance cannot have nucleon-kaon (two-particle) structure, but
rather is a genuine five-quark system or perhaps a more exotic
structure.

The semi-quantitative knowledge obtained with the schematic model can
also be used for molecules due to the use of dimensionless parameters.
For molecules we choose as typical scale units the mass of the $^4$He
atom and the range $b=10$ \AA. The $^3$He+$^4$He dimer has then $2m
b^2 |E_R| / \hbar^2 \approx 0.75$ \cite{nie98}. We can then expect
that, except for center-of-mass effects, the trimers
$^3$He+$^4$He+$^4$He and $^4$He+$^3$He+$^3$He should both have an
unbound state with dimensionless energy of about 1.50 when the
interaction between the two identical particles is neglected. However,
including these interactions of dimensionless energies 0.02 for
$^4$He+$^4$He and 5.2 for $^3$He+$^3$He, we arrive at trimers
$^3$He+$^4$He+$^4$He and $^4$He+$^3$He+$^3$He respectively bound and
unbound as observed.

\section{Summary and conclusions}

The two-body interactions are assumed to determine completely the
three-body structure including the continuum properties.  Close to the
threshold of binding, slightly above or below, the structure of the
three-body system mostly depends on the two-body low-energy scattering
properties. These properties are periodically repeated with increasing
strengths of these attractive potentials corresponding to one or many
bound states in each of the investigated channels.  Therefore it is
sufficient to study unbound or weakly bound two- and three-body
systems.  

The low-energy scattering properties are determined by the phase
shifts and reflected in the poles of the $S$-matrix, i.e.  by the
resonances and virtual states. Thus, changing the two-body
interactions simultaneously change the energies of both two- and
three-body resonances.  The relative changes of these energy
observables are expected to be intuitively easier to understand than by
using the connection via the two-body interaction strengths. The
strategy is then to vary the two-body resonances and virtual states
and study the changes of the corresponding quantities in the
three-body system.

We employ the efficient and well tested method of complex scaled
hyperspherical adiabatic expansion. This method is first briefly
sketched to define the notation. Then we define a schematic model with
an infinitely heavy core and two mutually non-interacting particles. We
prove mathematically for spin-zero particles that our formulation
provides a pole in the three-body $S$-matrix if and only if the
complex energy is equal to the sum of two complex energies each
corresponding to poles of different two-body $S$-matrices for the two
particle-core subsystems. The generalization to non-zero spins are
formulated and shown to involve diagonalization of simple
block-diagonal Hamiltonians.

The general properties are demonstrated numerically in schematic
examples involving both resonances and virtual states. We then
investigate the sizes and trends resulting from a finite core mass,
non-zero interaction between the light particles and the coupling of
different orbital angular momenta.  We use dimensionless units to
allow easy application on physics systems of different scales.  Each
of these effects can be substantial but the structure of the states
can be uniquely traced back to the origin in the structure of the
schematic model.  Borromean systems can arise with only two
interactions from center-of-mass effects and a favorable coupling of
two angular momenta.

For further numerical illustration we use essentially realistic
nuclear three-body systems, $^6$He, $^6$Li, and $^6$Be, consisting of
two nucleons and an $\alpha$-particle to trace back their measured
spectra to our bare schematic model. The effects of isospin symmetry
and the mixing short-range and Coulomb interactions are then seen. For
completeness we also present unpublished and fully realistic
calculations of $^6$Li and $^6$Be. The origin of the structure is
still apparent, but to get accurate energies and wave functions we
must include effects of spin-orbit couplings and the Pauli principle.
Three-body Thomas-Ehrman shifts can then be studied for these isobaric
analog states. Finally the effects of non-zero core spin is investigated
for the Borromean nucleus $^{17}$Ne consisting of $^{15}$O and two
protons.

In the last section we explain, and illustrate by examples, how to
make qualitative estimates of the three-body energies and their
structure from the two-body properties of the subsystems. First we
test by known examples of Borromean halo nuclei, hypertriton,
molecular helium clusters and the recently highlighted more
speculative pentaquark.

In conclusion, we have demonstrated the strong correlation between
two- and three-body resonances. The three-body energy and structure
can be traced back to the properties of an infinitely heavy core and
two non-interacting light particles. Substantial changes are often
needed to arrive at accurate and realistic properties but the generic
origin is apparent and revealing both structures and energies.

\vspace*{0.7cm}

\appendix

\section{Useful integrals}

The three-body $S$-matrix for the schematic model in Eq.(\ref{eq17})
is obtained through the definition in Eq.(\ref{eq16}) by inserting
Eq.(\ref{eq15}) into Eq.(\ref{eq12}). Then the following integrals
are needed:

\begin{eqnarray}
& & 
\int_0^{\pi/2} d\alpha (\sin \alpha)^{\ell_x+2} (\cos \alpha)^{\ell_y+2}
P_n^{(\ell_x+\frac{1}{2},\ell_y+\frac{1}{2})}(\cos 2\alpha)
\nonumber \\ & & \hspace*{3cm}
\times h_{\ell_x}^{(1)}(k_x x) h_{\ell_y}^{(1)}(k_y y) =   
 \\ & & \hspace*{-5mm}
(-1)^n \pi (\sin \alpha_\kappa)^{\ell_x} (\cos \alpha_\kappa)^{\ell_y}
P_n^{(\ell_x+\frac{1}{2},\ell_y+\frac{1}{2})}(\cos 2\alpha_\kappa)
\frac{H_{K+2}^{(1)}(\kappa\rho)}{(\kappa \rho)^2} \nonumber
\end{eqnarray}

\begin{eqnarray}
& & 
\int_0^{\pi/2} d\alpha (\sin \alpha)^{\ell_x+2} (\cos \alpha)^{\ell_y+2}
P_n^{(\ell_x+\frac{1}{2},\ell_y+\frac{1}{2})}(\cos 2\alpha)
\nonumber \\ & & \hspace*{2cm}
\times h_{\ell_x}^{(2)}(k_x x) h_{\ell_y}^{(2)}(k_y y) =0 
\end{eqnarray}

\begin{eqnarray}
& & 
\int_0^{\pi/2} d\alpha (\sin \alpha)^{\ell_x+2} (\cos \alpha)^{\ell_y+2}
P_n^{(\ell_x+\frac{1}{2},\ell_y+\frac{1}{2})}(\cos 2\alpha)
\nonumber \\ & & \hspace*{4cm}
\times h_{\ell_x}^{(1)}(k_x x) h_{\ell_y}^{(2)}(k_y y) =    
\nonumber \\ & &
\int_0^{\pi/2} d\alpha (\sin \alpha)^{\ell_x+2} (\cos \alpha)^{\ell_y+2}
P_n^{(\ell_x+\frac{1}{2},\ell_y+\frac{1}{2})}(\cos 2\alpha)
\nonumber \\ & & \hspace*{2cm}
\times h_{\ell_x}^{(2)}(k_x x) h_{\ell_y}^{(1)}(k_y y) 
\stackrel{\rho \rightarrow \infty}{\longrightarrow}   
 \\ & & \hspace*{-5mm}
(-1)^n \pi (\sin \alpha_\kappa)^{\ell_x} (\cos \alpha_\kappa)^{\ell_y}
P_n^{(\ell_x+\frac{1}{2},\ell_y+\frac{1}{2})}(\cos 2\alpha_\kappa)
\frac{H_{K+2}^{(2)}(\kappa\rho)}{2 (\kappa \rho)^2} \nonumber
\end{eqnarray}


\end{document}